# Zinc Sulphide Two-Dimensional Photonic Crystal Overlayer for Enhanced Extraction of Light from a Micro Cavity Light Emitting Diode

Michael A. Mastro[1], Chul Soo Kim[2], Mijin Kim[2], Josh Caldwell[1], Ron T. Holm[1], Igor Vurgaftman[2], Jihyun Kim[3], Charles R. Eddy Jr.[1] and, Jerry R. Meyer[2]

[1]U.S. Naval Research Laboratory, Electronics Science & Technology Division, 4555 Overlook Ave., SW, Washington, D.C. 20375, U.S.A.

[2]U.S. Naval Research Laboratory, Optical Science Division, 4555 Overlook Ave., SW, Washington, D.C. 20375, U.S.A.

[3]Department of Chemical and Biological Engineering, Korea University, Seoul, Korea

**Abstract**
A two-dimensional (2D) ZnS photonic crystal was deposited on the surface of a one-dimensional (1D) III-nitride micro cavity light emitting diode (LED), to intermix the light extraction features of both structures (1D+2D). The deposition of an ideal micro-cavity optical thickness of ≈ λ/2 is impractical for III-nitride LEDs, and in realistic multi-mode devices a large fraction of the light is lost to internal refraction as guided light. Therefore, a 2D photonic crystal on the surface of the LED was used to diffract and thus redirect this guided light out of the semiconductor over several hundred microns. Additionally, the employment of a post-epitaxy ZnS 2D photonic crystal avoided the typical etching into the GaN:Mg contact layer, a procedure which can cause damage to the near surface.

## 1. Introduction

The widespread adoption of III-nitride based solid-state lighting into a multitude of applications spaces, particularly general illumination, depends on an order of magnitude increase in the light output per dollar cost – or similar metric.[1,2] The cost factor will only decrease slowly over time, owing primarily to a shift to processing larger wafers. The total light output of commercial devices can be roughly delineated into internal and external efficiencies. Great progress has been made in blue InGaN-based optoelectronics, with internal efficiencies now approaching 60%, although this lags the nearly 100% internal efficiency of AlGaInP light-emitting devices.[3-6]

Nevertheless, all light emitting diodes (LEDs) suffer from poor extraction efficiency due to the large index of refraction contrast between the semiconductor and air. Light generated internal to the semiconductor cannot escape the material at an angle greater than the critical angle, $\theta_c = \sin^{-1}(n_{air}/n_s)$, where $n_s$ is the index of refraction of the semiconductor.[7] Light propagating at any angle further from the surface normal may not be extracted because it is subjected to total internal reflection. Thus, the external extraction efficiency from the top and bottom of an isotropic planar LED is quite low, with $2\eta = 8\%$ for an index of $n_{GaN} = 2.5$. This can be somewhat mitigated by a polymeric hemispheric dome (refractive index = 1.5), which increases the effective light cone at a cost of decreased radiance.[8,9]

Emission from a dipole placed in a cavity will experience multiple reflections.[10] A properly designed cavity will create interference effects that alter the internal angular light distribution. Light will constructively interfere in a thin cavity, since the difference in optical length for light traveling perpendicular to the surface and light traveling at the edge of the escape cone can be diminished to much less than a wavelength, as detailed by Benisty et al.[11]

Enclosing a functional LED in a micro cavity is difficult in III-nitride materials.[12,13] The standard procedure entails first growing 3 to 4 microns of buffer layer to reduce the dislocation density to a reasonable level ($10^8 cm^{-3}$).[4] Mastro et al. previously described a technique to produce a high-reflectance distributed Bragg reflector (DBR) with a limited number of bi-layers on a large-area Si substrate.[14,15] Related work showed this thin DBR to be advantageous for creating micro cavity LEDs (MCLEDs).[16] Nevertheless, even in these relatively thin cavities, multiple modes exist and a significant fraction of the light is trapped internal to the semiconductor.[17-20] This provided the motivation to introduce a 2D photonic crystal on the surface of the semiconductor to scatter this guided light. This article details a unique post-growth ZnS 2D photonic crystal, that is used to extract a portion of the light normally lost as guided modes internal to the one-dimensional (1D) MCLED.

## 2. Experimental

Micro cavity LEDs were grown directly onto Si(111) substrates by metal organic chemical vapor deposition. An alternating AlN/GaN sequence functions optically as a high-reflectance DBR and structurally as a strain-compensating superlattice. The growth was carried out in a modified vertical-impinging-flow chemical-vapor-deposition reactor. Two-inch Si wafers were cleaned via a modified Radio Corporation of America process followed by an in situ $H_2$ bake. An Al seed layer was deposited prior to the onset of $NH_3$ flow to protect the Si surface from nitridization. The AlN/GaN superlattice was deposited at 1050°C and 50 Torr. The reflection band-width for the DBR structure can be tuned to any point in the near-UV or visible by adjusting the thicknesses of the AlN and GaN layers in the DBR. In this article, a reflection band-width centered in the blue-green at 495 nm was obtained by alternating 59.9 nm AlN and 50.3 nm GaN layers. A measured reflectance from this DBR structure of up to 96.3% was previously reported.[14,15] A GaN SQW emitter was deposited at 1020°C at 150 Torr directly on the DBR, as well as on a control GaN-on-Si structure. The n- and p-type doping was accomplished with disilane and $Cp_2Mg$, respectively.

A backside contact was made with a Ni(15nm)/Au(150nm) bi-layer by e-beam evaporation. A similar bi-layer structure was processed on the top surface into an array of 200-m diameter contacts with 100-m spacing. Then the PC patterns were e-beam lithographically formed to eight identical 100 m x 100 m boxes with a triangular 2D photonic crystal pattern in proximity to a top contact. The positive e-beam resist (ZEP520A, Zeon Chemicals Co.) with 350 nm of thickness was spin coated on the surface of the top GaN layer. E-beam lithography is performed at 25 kV beam voltage using a Raith 150 e-beam lithography system. After e-beam lithography, PC patterns were developed. ZnS (75 mn) was then deposited by e-beam evaporator followed by lift-off step. The triangular symmetry PC patterns were formed by high index ZnS film ($n$ = 2.36) and lattice period was chosen to be 170 nm, designed to match the UV emission of the underlying LED.

A constant current of 20 mA was injected through the vertical LED using a Keithley 2400 Source Measurement Unit. The electroluminescence (EL) spectrum was collected in the 100X (0.7 NA) objective of a Mitutoyo microscope and the spectral output was directed via an optical fiber into an Ocean Optics USB2000 spectrometer with an effective spectral range of 200 to 850 nm.

## 3. Results

A cross-section of the MCLED structure is discernable in the electron micrograph of Fig. 1. This sample consisted of a GaN single quantum well (SQW) LED, that was deposited on a 7x GaN/AlN DBR on a Si substrate. The LED comprised a 287-nm GaN:Si n-clad, a (10-nm / 10-nm / 10-nm) AlGaN:Si / GaN-undoped / AlGaN:Mg SQW, and a 144-nm GaN:Mg p-clad. The alternating sequence in the DBR was designed to absorb strain and filter dislocations generated at the GaN/Si interface.[15,21] The triangular ordering of the ZnS 2D photonic crystal is also shown in Fig. 1.

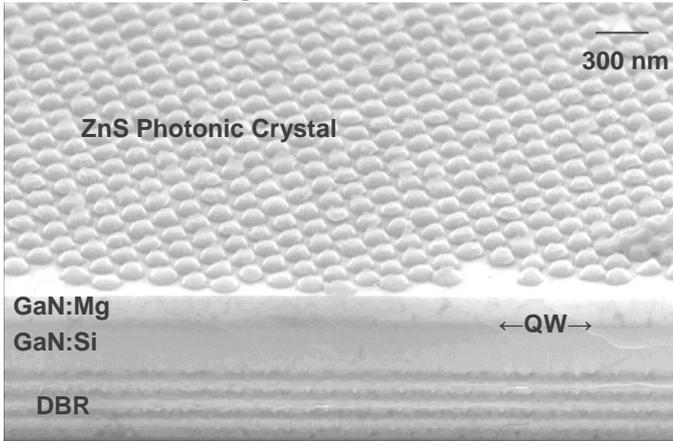

Figure 1 Cross-section electron micrograph of a SQW LED embedded in a micro cavity with bottom Bragg reflector. The inclined view shows the ZnS 2D photonic crystal on the surface of the LED.

Figure 2 shows optical pictures of the light extraction from LEDs with (top row) and without (bottom row) the photonic crystal. The light generated under the Ni/Au contact cannot penetrate the thick metal used in this simplified contact scheme. The dimensions of a micro-cavity limits the number of modes that exist and, thus, creates a sharp contrast between modes escaping perpendicular to the surface and modes trapped internal to the semiconductor. This is clearly seen as a distinctive ring of emission is near the contact edge. This light emitting ring defines an effective current spreading length in the top p-type layers at 20 mA of drive current.

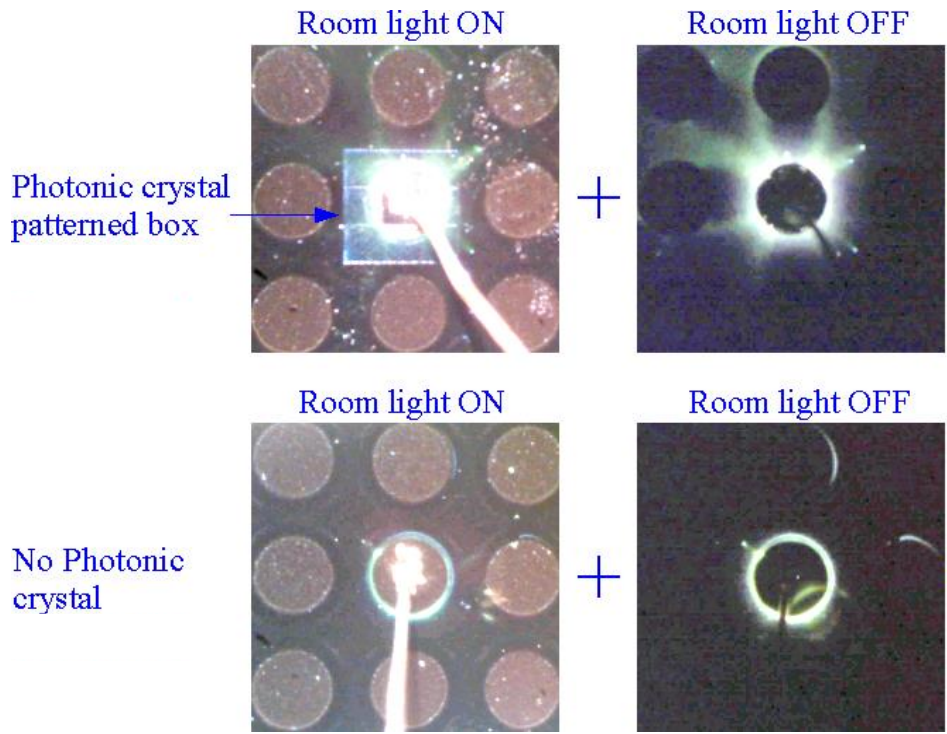

Figure 2 Light emission at drive currents of 20 mA. In the sample without the photonic crystal (bottom row), emission is observed only around the edge of the thick circular contact. In contrast, emission is clearly enhanced in the LED with the photonic crystal (top row) throughout the patterned region (seen as a box in the upper left picture). The eight 100 μm x 100 μm photonic crystal boxes frame the periphery of a 300 micron square, with a common central contact disc.

The top row of Fig. 2 reveals that the LEDs patterned with the photonic crystal display a substantial enhancement of the light extraction in regions spatially removed from the contact but still within the photonic crystal box. On average, the electroluminescence measurements indicated an overall 45% increase in light extraction with the addition of the ZnS photonic crystal. This enhancement is particularly striking away from the contact edge where there is essentially zero light emission without the photonic crystal.

Direct scaling of this 1D+2D photonic crystal structure is possible to extract the wavelength of interest. An alternate photonic crystal and underlying MCLED was fabricated with a structure designed to enhance the extraction of light with a wavelength near 420 nm. In this case, the ZnS lattice period was set to 193nm and the DBR was composed of 54.5 nm AlN and 45.2 nm GaN layers.

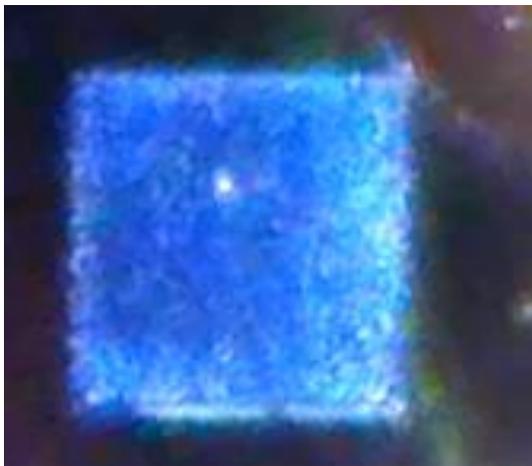

Figure 3. Light emission from a MCLED with a photonic crystal (large blue square) designed to extract emission in the blue regime at 420 nm. A section of the large area metal contact is observable in the upper right corner.

### 4. Discussion

To place this work into the context, the micro-cavity regime, described by Benisty et al. [11], does not apply to traditional 3-micron thick GaN LEDs on sapphire, where several Fabry-Perot oscillations are thus seen either across the 2-4 kT wide emission spectrum or from the presence of multiple angular lobes at a single wavelength. Conversely, the present thin MCLED (1 micron) does meet this criterion, and thus will alter the internal angular power distribution of an emitter and can force the preferred propagation direction of photons into the light extraction cone. Previous work by Mastro et al. showed that a micro-cavity could enhance or suppress the UV, blue, or green emission of a GaN SQW.[14-16] The micro-cavity was simplified in this work by relying on the air / semiconductor interface to act as the top mirror instead of an air/metal/GaN or air/DBR structure.

Light exiting into air directly from ZnS or GaN will experience a reflectance of only approximately 20%. Still, micro-cavity effects will exist in this design albeit with a much lower finesse in the cavity. Additionally, the lack of a current spreading metal and the spreading resistance in the GaN p-cladding layer prevents high current injection and, hence, high radiative emission rate away from the contact edge as seen in the bottom row of Fig. 2. Nevertheless, this design without a thin top metal was purposely chosen to isolate the ability of the 2D ZnS photonic crystal to extract light normally lost as a guided mode.

Zinc sulfide was selected for the photonic crystal, in part, because of its close index-of-refraction (n=2.36) match with GaN, which avoids reflection at the interface. Zinc sulfide also has a wide bandgap (3.6 eV), which allows for transparency in the UV and visible regimes.

It is well known that etching a photonic crystal into the p-type GaN layer does enhance the light extraction from a III-nitride LED. [22-31] As in the present work, the surface Bragg grating diffracts the internally trapped light. Unfortunately, processing of p-type GaN is notorious for introducing a defective n-type surface layer, which can hamper current injection and introduce mid-gap states that absorb light emission.[3,4] The process outlined in this article avoids any etching the p-type GaN layer.

The primary design of the lattice of ZnS cylinders is to diffract modes that would have otherwise been internally guided in the cavity. An extension of this work is underway to surround the ZnS cylinders with a thin layer of metal on the GaN surface to improve current spreading as well as increase the reflectance of the top mirror of the micro-cavity. Beyond this, the hemispheric top of the ZnS cylinder is intriguing and work is underway to study these as nanolenses on the surface of a LED.

## 5. Summary

Extraction efficiency from a III-nitride quantum well was enhanced by embedding the emission layer in a 1D micro cavity and also forming a ZnS 2D photonic crystal on the surface of the III-nitride LED. This unique design couples the enhancements of both structures (1D+2D), while relaxing the structural requirements for enhancement of the individual components.

## Acknowledgements

Research at the Naval Research Lab is partially supported by the Office of Naval Research and Office of Naval Research–Global (N00014-07-1-4035). Research at Korea University was supported by BK21 program.

## References

1. Gallium Nitride 2005 - Technology Status, Applications, Market Forecasts, Strategies Unlimited Report (2005)
2. Light Emitting Diodes (LEDs) for General Illumination Update 2002, DOE OIDA Roadmap (2002)
3. M. Henini, M Razeghi, Optoelectronic Devices: III Nitrides, Elsevier Science, Berlin (2005)
4. S. Nakamura, G. Fasol, S.J. Pearton, The Blue Laser Diode: The Complete Story, Springer, Berlin (2000)
5. J. J. Wierer, D. A. Steigerwald, M. R. Krames, J. J. O'Shea, M. J. Ludowise, G. Christenson, Y.-C. Shen, C. Lowery, P. S. Martin, S. Subramanya, W. Götz, N. F. Gardner, R. S. Kern, S. A. Stockman, Appl. Phys. Lett. **78**, 3379 (2001).
6. P. Altieri, A. Jaeger, R. Windisch, N. Linder, P. Stauss, R. Oberschmid, K. Streubel, J. Appl. Phys. **98** (2005) 086101
7. W. Lukosz, R. E. Kunz, *J. Opt. Soc. Amer.*, **67** (1977) 1607
8. C. Weisbuch, H. Benisty, R. Houdre, J. Lumin, **85** (2000) 271
9. M.G. Craford, G.B. Stringfellow, M.G. Crawford (Eds.), High Brightness Light Emitting Diodes, Academic Press, San Diego, 47, 1997
10. M. A. Mastro, R. T. Holm, N. D. Bassim, C. R. Eddy, Jr., R. L. Henry, M. E. Twigg, A. Rosenberg, Jpn. J. Appl. Phys, **45** (2006) L814
11. H. Benisty, H. De Neve, C. Weisbuch, IEEE J. Quantum Electron. **34** (1998) 1612
12. D. Delbeke, R. Bockstaele, P. Bienstman, R. Baets, H. Bensity, IEEE J. Quantum Electron. **8** (2002) 189
13. E. Schubert, N. Hunt, M. Micovic, R. Malik, D. Sivco, A. Cho, G. Zydzik, Science, **265** (994) 943
14. M.A. Mastro, R.T. Holm, N.D. Bassim, C.R. Eddy Jr., D.K. Gaskill, R.L. Henry, M.E. Twigg, Appl. Phys. Lett. **87** (2005) 241103
15. M.A. Mastro, R.T. Holm, N.D. Bassim, D.K. Gaskill, J.C. Culbertson, M. Fatemi, C.R. Eddy Jr., R.L. Henry, M.E. Twigg, J. Vac. Sci and Tech. A **24** (2006) 1631
16. M.A. Mastro, J. D. Caldwell, R. T. Holm, R. L. Henry, C. R. Eddy Jr., Adv. Mat. **20**-1, 115 (2008)
17. H. Benisty, H. De Neve, C. Weisbuch, J. Quantum Electron., **34** (1998) 1632
18. H. Benisty, R. Stanley, M. Mayer, J. Opt. Soc. Amer. A, **15** (1998) 1192
19. J. Dorsaz, J.-F. Carlin, S. Gradecak, M. Ilegems, J. Appl. Phys. **97** (2005) 084505
20. J.-F. Carlin, C. Zellweger, J. Dorsaz, S. Nicolay, G. Christmann, E. Feltin, R. Butt, N. Grandjean, Phys. Stat. Sol. (b) **242** (2005) 2326
21. M.A. Mastro, C.R. Eddy Jr., D.K. Gaskill, N.D. Bassim, J. Casey, A. Rosenberg, R.T. Holm, R.L. Henry, M.E. Twigg, J. Crystal Growth, **287** (2006) 610
22. D. Ochoa, R. Houdré, R. P. Stanley, C. Dill, U. Oesterle, M. Ilegems, J. Appl. Phys. **85** (2006) 2994
23. K. Orita, S. Tamura, T. Takizawa, T. Ueda, M. Yuri, S. Takigawa, D. Ueda, Jap. J. Appl. Phys. **43** (2004) 5809

24. J. J. Wierer, M. R. Krames, J. E. Epler, N. F. Gardner, M. G. Craford, Appl. Phys. Lett. **84** (2004) 3885.
25. A. David, H. Benisty, C. Weisbuch, "Optimization of Light-Diffracting Photonic-Crystals for High Extraction Efficiency LEDs," J. Display Technol. **3** (2007) 133
26. A. David, C. Meier, R. Sharma, F. S. Diana, S. P. DenBaars, E. Hu, S. Nakamura, C. Weisbuch, H. Benisty, Appl. Phys. Lett. **87** (2005) 101107
27. A. David, T. Fujii, R. Sharma, K. McGroddy, S. P. DenBaars, E. Hu, S. Nakamura, C. Weisbuch, H. Benisty, Appl. Phys. Lett. **88** (2006) 061124
28. J. J. Wierer, D. A. Steigerwald, M. R. Krames, J. J. O'Shea, M. J. Ludowise, G. Christenson, Y.-C. Shen, C. Lowery, P. S. Martin S. Subramanya, W. Götz, N. F. Gardner, R. S. Kern, S. A. Stockman, Appl. Phys. Lett. **78** (2001) 3379.
29. M.A. Mastro, J. A. Freitas, Jr., O. Glembocki, C.R. Eddy, Jr., R. T. Holm, R.L. Henry, J. Caldwell, R.W. Rendell, F. Kub, J-H. Kim, Nanotechnology, **18** (2007) 265401
30. H. Yokoyama, Science **256** (1992) 66
31. M.A. Mastro, D.V. Tsvetkov, A.I. Pechnikov, V.A. Soukhoveev, G.H. Gainer, A. Usikov, V. Dmitriev, B.Luo, F. Ren, K.H. Baik, S.J. Pearton, Mat. Res. Soc. Proc. **764** (2003) C2.2